\documentclass[aps,prl,reprint,twocolumn,showpacs,floatfix,superscriptaddress]{revtex4-2}

\usepackage{amsmath,amssymb,amstext} 
\usepackage{verbatim}
\usepackage{graphicx}
\usepackage{color}
\usepackage[dvipsnames]{xcolor}
\usepackage{tikz}
\usepackage[colorlinks=true,citecolor=blue,linkcolor=blue,urlcolor=blue]{hyperref}
\usepackage{braket, enumitem,}
\usepackage[normalem]{ulem}

\def\be{\begin{equation}}
\def\ee{\end{equation}}
\def\bea{\begin{eqnarray}}          
\def\eea{\end{eqnarray}}
\def\bi{\begin{itemize}}
\def\ei{\end{itemize}}

\usepackage{longtable}
\usepackage{epsfig}
\usepackage{dcolumn}
\usepackage{bm}
\usepackage{babel}
\usepackage{natbib}

\makeatother

\begin{document}

\title{
Non-adiabatic dynamics across a first order quantum phase transition: \\
Quantized bubble nucleation
}

\author{Aritra Sinha}
\email{aritrasinha98@gmail.com}
\affiliation{Institute of Theoretical Physics, Jagiellonian University,   \L{}ojasiewicza 11, 30-348 Krak\'ow, Poland }
            
\author{Titas Chanda}
\email{titas.chanda@uj.edu.pl}
\affiliation{Institute of Theoretical Physics, Jagiellonian University,   \L{}ojasiewicza 11, 30-348 Krak\'ow, Poland }

\author{Jacek Dziarmaga}
\email{dziarmaga@th.if.uj.edu.pl}
\affiliation{Institute of Theoretical Physics, Jagiellonian University,   \L{}ojasiewicza 11, 30-348 Krak\'ow, Poland }

\date{\today}

\begin{abstract}
Metastability is a quintessential feature of first order quantum phase transitions, which is lost either by dynamical instability or by nucleating bubbles of a true vacuum through quantum tunneling. By considering a drive across the first order quantum phase transition in the quantum Ising chain in the presence of both transverse and longitudinal fields, we reveal multiple regions in the parameter space where the initial metastable state loses its metastability  in successive stages. The mechanism responsible is found to be semi-degenerate resonant tunnelings to states with specific bubble sizes. We show that such dynamics of \textit{quantized} bubble nucleations can be understood in terms of Landau-Zener transitions, which provide quantitative predictions of nucleation probabilities for different bubble sizes. 
\end{abstract}

\maketitle


\textbf{Introduction.--}
Non-equilibrium aspects of many-body quantum systems are at the heart of understanding the fundamentals of statistical and condensed matter physics as well as of quantum field theory~\cite{Srednicki1994, Calabrese2004, Calabrese2005, Rigol2008, Calabrese2009, Polkovnikov2011, Eisert2015, DAlessio2016,rossini2021}. On the theoretical front, the analysis of the dynamics of non-integrable systems have soared drastically during the last two decades due to the advancement and development of efficient numerical tools like various tensor networks methods \cite{Schollwock2011, Orus2014, Paeckel2019}. Moreover, with the recent breakthroughs in quantum simulations \cite{Feynman1982, Johnson2014, Cirac2012, Bloch2012, Blatt2012, AspuruGuzik2012, Gross2017}, studying the non-equilibrium features of complex quantum systems on table-top experiments has become a reality, especially in the substrates like cold-atoms on optical lattices \cite{Gring2012, Yan2013, Paz2013, Schreiber2015, Bernien2017, Lukin18} or trapped ions \cite{Jurcevic2014, Richerme2014, Bohnet2016, Smith2016, Martinez2016, Neyenhuis2016, Jurcevic2017, Monroe2019}.

One promising avenue of work in this ubiquitous facet of fundamental physics has been to investigate non-adiabatic excitations due to quenches across a continuous quantum phase transition \cite{Sachdev2009, Dutta2015} under the unifying framework of the quantum version \cite{Damski2005, Zurek2005, Dziarmaga2005, Polkovnikov2005, Dziarmaga2010, Polkovnikov2011} of the classic Kibble-Zurek (KZ) mechanism \cite{K-a, K-b, K-c, Z-a, Z-b, Z-c}. However, the  question that has been asked less frequently is regarding the consequence of a slow drive across a first order quantum phase transition (FOQPT) \cite{Pfleiderer2005, Vojta2003} and whether it is possible to find any similar universal dynamical features akin to quantum KZ theory.

FOQPTs exhibit metastability on a drive across the transition, i.e., the system tends to persist in the local minimum due to the presence of a potential barrier. In the traditional language of continuous field theory, the state gets stuck in a false vacuum, that is stable against small fluctuations, and cannot tunnel to the true vacuum easily. However, on a dynamical quench across FOQPT, the false vacuum may become dynamically unstable and the true vacuum may develop due to the disappearance of the potential barrier far beyond the FOQPT point. Under such scenarios, several recent studies reported KZ scaling laws for the dynamics across certain first order phase transitions -- both classical as well as quantum \cite{Swislocki2012, Panagopoulos2015, Pelissetto2017, Coulamy2017, Liang2017,  Shimizu2018, Qiu2020, Andrea2018, Andrea2020}.
Another more generic mechanism, through which such metastability can evaporate, is the continual creation of bubbles of the true vacuum driven by the quantum fluctuations inside the false vacuum. The aim of the present letter is to thoroughly investigate breakdown of metastability by the nucleation of bubbles in a many-body quantum setting -- going beyond the paradigm of dynamical instability and the corresponding KZ mechanism.

We consider the generic one-dimensional (1D) quantum Ising chain in the presence of both transverse and longitudinal fields. The model possesses a FOQPT between two ferromagnetic phases of opposite orientations driven by the longitudinal field. On slow tuning of the longitudinal field across the FOQPT line, we detect a multitude of special (resonant) points/regions where the nucleation of bubbles of the true vacuum inside the metastable false vacuum becomes energetically favorable. Moreover, these tunneling processes are \textit{quantized} in the sense that only a specific size of bubbles, pertaining to a specific perturbative order, can nucleate around the corresponding resonant value of the longitudinal field. We provide accurate quantitative explanations of these non-adiabatic changes by means of the archetypal Landau-Zener (LZ) theory \cite{Landau65, Zener32, Stueckelberg32, Majorana32}.

\textbf{Model.--} 
The quantum Ising model in the presence of a transverse field in 1D is one of the prototypical models used for several decades to understand the quantum phase transition at zero temperature \cite{Sachdev2009, Dutta2015}. In the presence of an additional longitudinal field the Hamiltonian reads
\bea
H=
-\sum_{n=1}^N
\left[  
 \sigma^z_n\sigma^z_{n+1} \ + h_{x}\sigma^x_n \ +
 h_{z}\sigma^z_n
\right],
\label{hamiltonian}  
\eea
where we assume transverse field $h_x>0$ for definiteness. Apart from having a rich phase diagram, this model, although being simple, has become a test-bed for fascinating equilibrium as well as out-of-equilibrium phenomena, like weak-thermalization \cite{Olexei2017}, dynamical confinement \cite{Kormos2016, Verdel2020, Karpov2020, Surace2020}, existence of quantum many-body scars \cite{Michailidis2020, James2019, Robinson2019}, or fracton dynamics \cite{Pai2020}. For the longitudinal field $h_{z} = 0$, this model has a continuous quantum phase transition at $h_{x} = 1$ separating the ferromagnetic phase ($h_{x}<1$) from the paramagnetic one ($h_{x}>1$). An FOQPT exists separating two ordered ferromagnetic phases along the so-called Ising line ($h_{z} = 0$). In this letter, we will mostly stay in the regime of small transverse field, $h_x\ll1$ ({although this is not a strict requirement}), where, with the exception of some special regions, it can be considered as a source of small quantum fluctuations in a classical Ising chain. 

\textbf{Linear ramp and special regions.--}  
To initiate, we prepare the system in the ground state of the Hamiltonian \eqref{hamiltonian} in one of the ordered phases ($h_x < 1$) and perform a slow ramp of the longitudinal field $h_{z}$ across the first order transition. We choose a protocol akin to what is usually used in studies of the KZ mechanism, where we ramp the field as
\bea
h_{z}(t) = h_z^{\text{in}} + \frac{t}{\tau_Q}.
\label{quench_protocol}  
\eea
We start at time $t=0$ when the initial state $\ket{\psi(t=0)} = \ket{\psi_{\text{in}}}$ is the ground state of the Hamiltonian \eqref{hamiltonian} with $h_{z}^{\text{in}} < 0$, and then ramp the field up to a final value $h_{z}^{\text{fin}} > 0$ in the opposite ordered phase. The total ramp time is proportional to $\tau_{Q}$. The dynamics is simulated by using time-dependent variational principle (TDVP) \cite{Haegeman2011, Koffel2012, Haegeman2016, Paeckel2019} based on matrix-product state (MPS) \cite{Schollwock2011, Orus2014} ansatz with open boundary condition. 

To start, we use average longitudinal magnetization 
$  
m_{z}  = \frac{1}{N}\sum_n \braket{\sigma^{z}_{n}},
$
and the longitudinal density of kinks $ \Gamma = \frac{1}{2}\left(1-\frac{1}{N-1}\sum_n \braket{\sigma_{n}^{z}\sigma_{n+1}^{z}}\right)$ as our \textit{bona fide} observables. Deep in the ferromagnetic phase, for $|h_z^{\text{in}}| \gg h_x$, the initial state has $m_z \approx -1$ and $\Gamma \approx 0$ as the state is highly polarized:
\be 
\ket{\psi_{\text{in}}} \approx \ket{\downarrow \downarrow \ldots \downarrow}.
\label{psiin}
\ee 
 
\begin{figure}
\includegraphics[width=\linewidth]{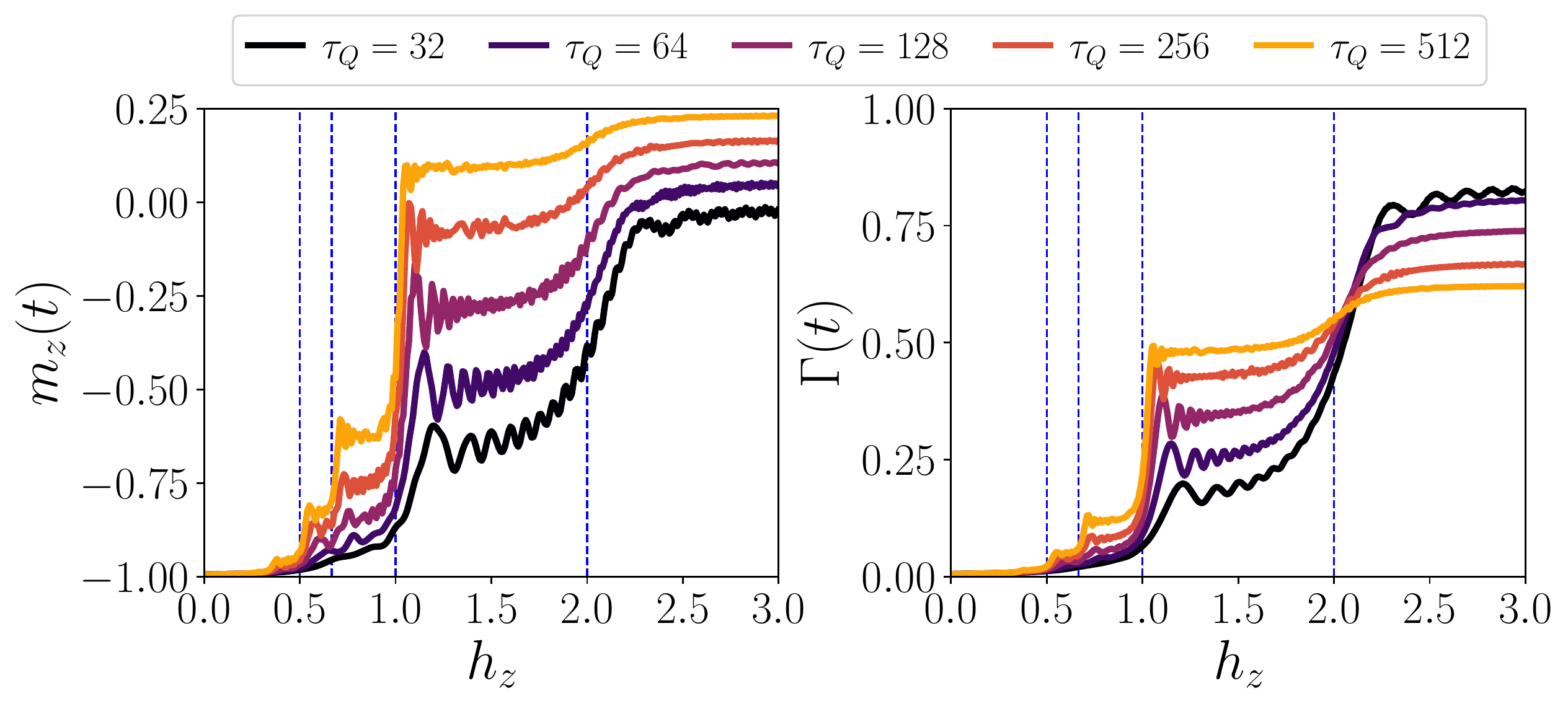}
\caption{
The average magnetization $m_z(t)$ (left) and the density of kinks $\Gamma(t)$ (right)
tracked dynamically by varying the parallel field $h_{z}$ according to Eq. \eqref{quench_protocol} for a wide range of quench times $\tau_{Q}$. Here we set $h_{x}=0.2$ and  $h^{\textrm{in}}_{z}= -4.0$.
}
\label{fig:ramp} 
\end{figure}

During the ramp, we observe that the initial state remains metastable against small quantum fluctuations driven by the transverse $h_x$ for a long time after crossing the FOQPT point. The crucial feature in this scenario is that the system departs from the metastable state at several \textit{special occasions} during the ramp, see Fig.~\ref{fig:ramp}. Deep into the positive ferromagnetic phase ($h_z > 0$), the average magnetization $m(t)$ and the density of kinks $\Gamma(t)$ get jolted up in several steps and finally saturates although with visible small oscillations. We shall show that these special regions exist around points where the initial metastable state is semi-degenerate with bubbles of the true vacuum. Below, we provide an heuristic explanation first.

Without quantum fluctuations, when the knob is set to $h_x=0$, any spin flips in the fully polarized initial state (\ref{psiin}) increases the ferromagnetic energy -- a domain or bubble of $n$ consecutive $\uparrow$-spins increases the ferromagnetic energy by $4$, regardless of its size. Overall, taking into account the local longitudinal fields, such a bubble of size $n$ changes the total energy by $4-2n h_z$, which becomes zero when
\bea
h_{z} = 2/n.
\label{quant_bubbles}
\eea
As a result, we have \textit{quantized} values of the longitudinal field $h_{z} = 2,1,2/3,1/2,...$, where the initial metastable state becomes degenerate with states having bubbles of sizes $n = 1,2,3,4,...$, respectively. These are the four major quantized values of the longitudinal magnetic field that correspond to the jolts clearly seen in Fig. \ref{fig:ramp}. 

\begin{figure}
\includegraphics[width=\columnwidth]{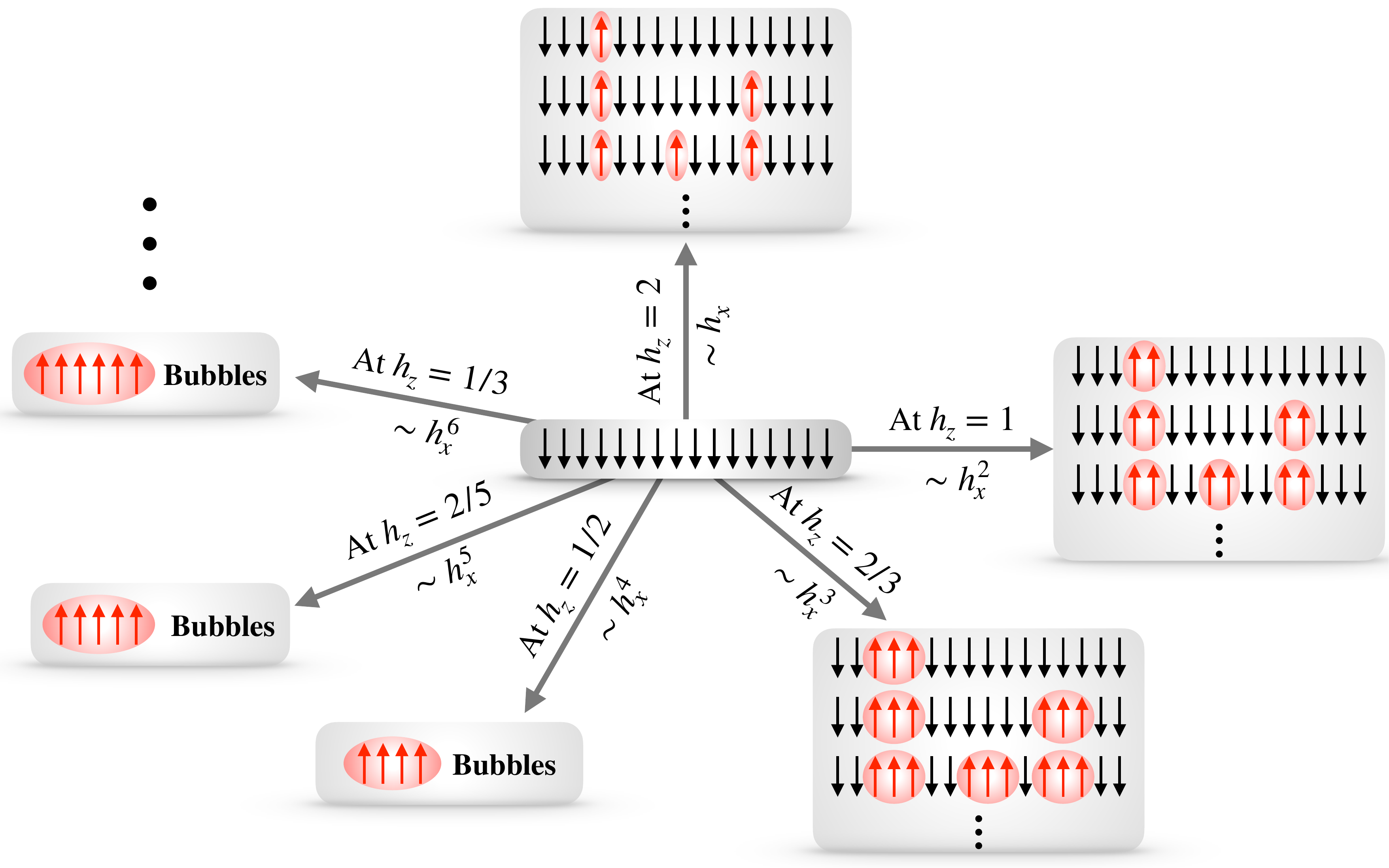}
\caption{
Diagram showing the possible nucleation processes and their perturbative orders in $h_x$.
At $h_z=2/n$ the metastable initial state, with all spins pointing down, is semi-degenerate with states containing size-$n$ bubbles with spins up. They are connected by an $n$-th order process in $h_x$. 
}
\label{schematic_1} 
\end{figure}

When $h_x>0$, operators $\sigma^z_n$ are no longer good quantum numbers. For a generic $h_z$, the initial state (\ref{psiin}), dressed with quantum fluctuations of second order in $h_x$, remains an approximate eigenstate. This is not the case at the special quantized values of $h_z$, where the initial state becomes \text{semi-}degenerate with bubbles of size $n$ -- connected by anticrossings -- and even a tiny $h_x$ is enough to mix them. Each of these \textit{resonant points} in $h_z$ corresponds to a particular order in perturbation theory with respect to $h_x$. For instance, only nucleations of bubble-size $1$ can happen near $h_{z} = 2$. As this requires a single spin to be flipped, the tunneling between the degenerate states is first order in $h_x$. In general, tunneling at $h_z=2/n$ between the polarized initial state and a state with an $n$-bubble is an $n$-th order process. For a schematic viewpoint see Fig.~\ref{schematic_1}. 

\textbf{Landau-Zener (LZ) nucleation theory.--} 
We begin with $n=1$ near $h_z=2$. For low density of flipped spins, we can consider flipping an isolated spin at site $j$:
\be 
\ket{ \downarrow\dots\downarrow \downarrow_j \downarrow\dots\downarrow }
\stackrel{h_x}{\longleftrightarrow} 
\ket{ \downarrow\dots\downarrow  \uparrow_j  \downarrow\dots\downarrow }.
\ee 
The tunneling is driven by the term $-h_x\sigma^x_j$. In the two dimensional subspace,
the Hamiltonian reads
\begin{equation}
H_{\text{eff}}^{(1)} = E_0(h_z)+\begin{bmatrix}
0 & -h_{x}\\
-h_{x} & 4-2h_{z} 
\label{heff_1bub}
\end{bmatrix},    
\end{equation}
where $E_0(h_z)=-(N-1)+Nh_z$ is the energy of the metastable state. With the linear ramp~\eqref{quench_protocol} this becomes the LZ problem \cite{Landau65, Zener32, Stueckelberg32, Majorana32} with an anticrossing when $h_z=2$. The LZ probability to flip the spin is
$p_1=1-\exp(-\pi\tau_{Q}h_{x}^2)$. Beyond the two-dimensional subspace, this formula is accurate only when $p_1\ll1$ or, equivalently, for fast quenches with $\pi\tau_{Q}h_{x}^2\ll1$. Otherwise, the density of flipped spins becomes large and we cannot consider flipping spin $j$ in isolation from flipping other spins.

\begin{figure}
\includegraphics[width=\linewidth]{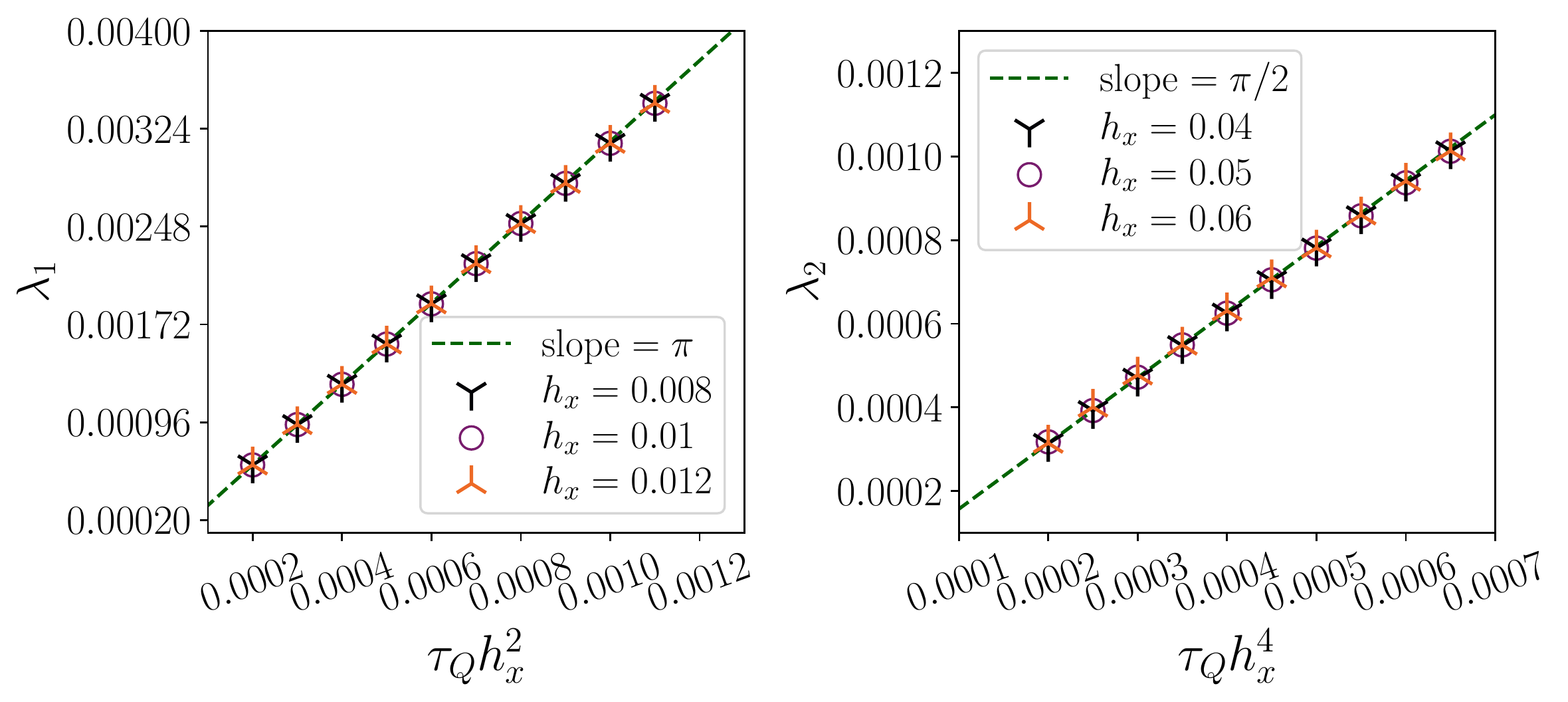}
\caption{The density of $1$-bubbles, $\lambda_{1}$ (left), and  $2$-bubbles, $\lambda_{2}$ (right), are shown as functions of scaled $\tau_{Q}$ for several strengths $h_x$ of quantum fluctuations. The different $h_x$ collapse to straight lines with slopes $\pi$ and $\pi/2$ for $n=1,2$, respectively. The collapse demonstrates the accuracy of the simple Landau-Zener theory for low density of nucleated bubbles.}
\label{fig:bub_gen} 
\end{figure}

More generally, bubbles of $n$ spins are nucleated near $h_z=2/n$. For a low total density of bubbles we can consider flipping $n$ consecutive spins $j,\dots,j+n-1$ by a $n$-th order process:
\small
\be 
\ket{ \downarrow\dots\downarrow \downarrow_j\dots\downarrow_{j+n-1} \downarrow\dots\downarrow } 
\stackrel{h_x^n}{\longleftrightarrow} 
\ket{ \downarrow\dots\downarrow \uparrow_j  \dots  \uparrow_{j+n-1} \downarrow\dots\downarrow }.
\ee
\normalsize
For such a process the effective Hamiltonian reads
\begin{equation}
H_{\text{eff}}^{(n)} \approx E_0(h_z)+
\begin{bmatrix}
0          & -c_n h_{x}^{n}\\
-c_n h_{x}^{n} & 4-2n h_{z} 
\label{heff_nbub}
\end{bmatrix}.    
\end{equation}
Here $c_n$ is a combinatorial factor.  In general, it can be derived for any order $n$ by treating the transverse field perturbatively and  obtaining the low-energy effective Hamiltonian through the Schrieffer-Wolff transformation ~\cite{Wolff1966}. For particular perturbative orders, we will concentrate on  $c_{1} = c_{2}=1$ and $c_{3}=81/64$ in this letter ~\cite{supple}. After the anticrossing at $h_z\approx 2/n$ the LZ probability to nucleate the $n$-bubble reads ~\cite{supple}
\be 
p_n=
1-\exp\left(-\frac{c_n^2}{n}\pi\tau_{Q}h_{x}^{2n}\right)
\approx
\frac{c_n^2}{n}\pi\tau_{Q}h_{x}^{2n}.
\label{linLZ}
\ee 
It is accurate for $\tau_{Q}h_{x}^{2n}\ll1$ only. In order to verify the LZ formula, we consider the density of $n$-bubbles:
\begin{equation}
    \lambda_{n} = \left\langle P^{\downarrow}_{i}  \left[\prod_{j=1}^{n} P^{\uparrow}_{i+j}\right]  P^{\downarrow}_{i+n+1}\right\rangle.
    \label{bub_density_eq}
\end{equation}
Here $P_{j}^{\uparrow,\downarrow}=\left(1\pm\sigma_{j}^{z}\right)/2$ is a projector onto spin-$\uparrow$($\downarrow)$ at site $j$ and $\braket{\dots}$ refers to averaging over all sites except for the ends of the chain to avoid boundary effects. In Fig. \ref{fig:bub_gen} we plot $\lambda_1$ and $\lambda_2$ obtained with TDVP as a function of $\tau_{Q}h_{x}^{2n}$ for several values of $h_x$ such that the low density condition, $\tau_{Q}h_{x}^{2n}\ll1$, is satisfied. Plots for different $h_x$ collapse to a straight line with a slope consistent with the simple LZ theory.

\textbf{Nucleation versus hopping.--}
In the second order perturbation in $h_x$, the $n$-bubble at sites $j,\dots,j+n-1$ can hop to the right/left by one lattice site. In order to hop to the right, spin $j+n$ can be flipped upwards followed by a downward flip of spin $j$, or the other way round. The net hopping rate is
$ 
\gamma = h_x^2/h_z.
$

The LZ formula cannot be taken for granted if the nucleated bubble can hop away before the LZ tunneling is completed
at time
$
t_{\rm LZ} \approx \sqrt{\tau_Q/2n} 
$
\cite{Damski2005} after the anticrossing at $h_z=2/n$. Therefore, the hopping should be irrelevant when $\gamma t_{\rm LZ}\ll 1$ or, equivalently,
\be 
\frac18 n\tau_Qh_x^4\ll 1.
\label{hopir}
\ee
For $1$-bubbles this condition is satisfied with a safe margin in their low density regime where $\pi\tau_Qh_x^2\ll1$. For $2$-bubbles it is identical with low density. However, for $3$-bubbles and bigger it is stronger than low density. For $3$-bubbles the hopping is a second order process while the LZ tunneling is formally a weaker third order effect. 

\begin{figure}
\includegraphics[width=0.8\linewidth]{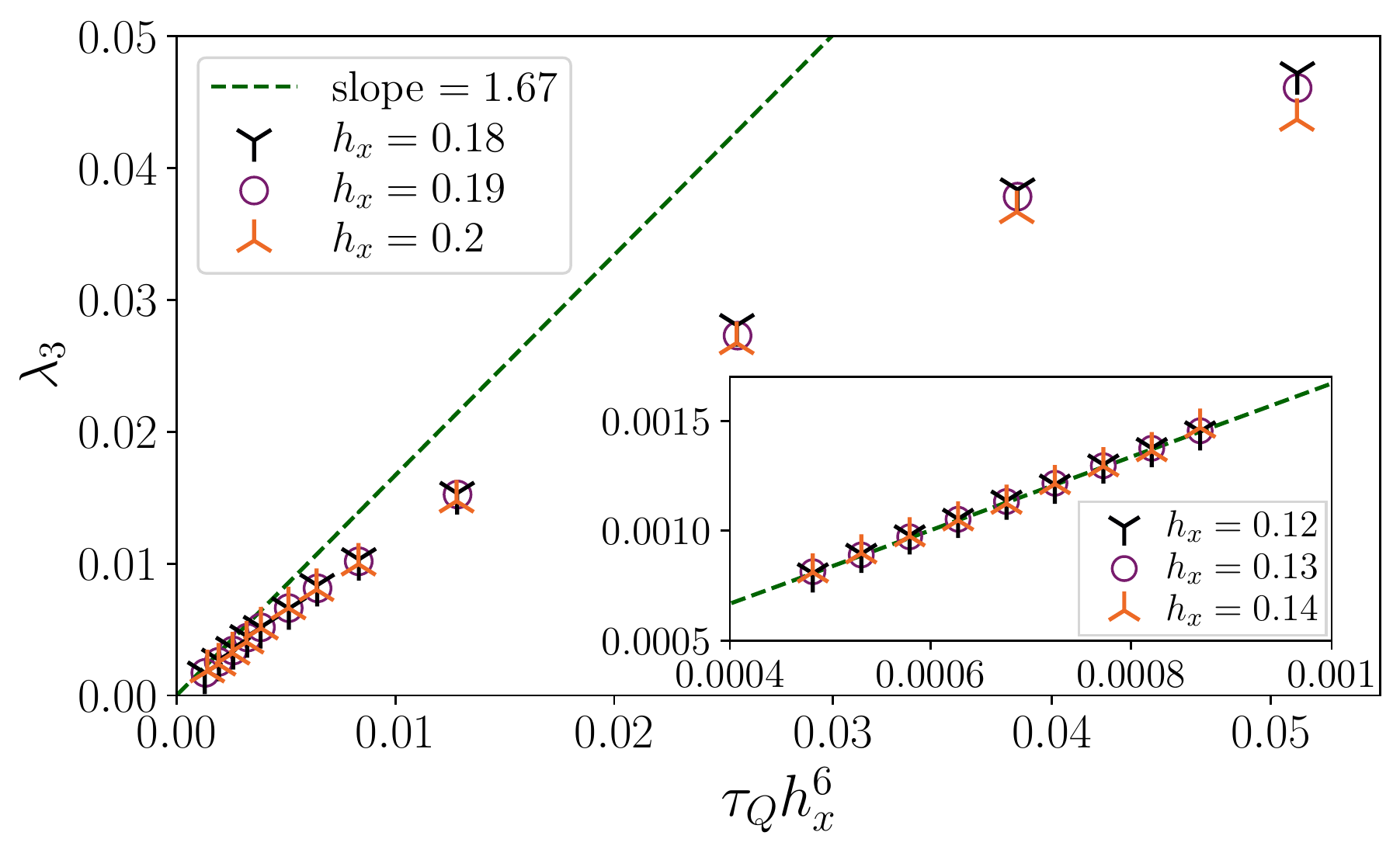}
\caption{
Density of $3$-bubbles, $\lambda_{3}$, as a function of the scaling variable $\tau_Qh_x^6$ for different values of $h_x$. Here $\tau_Qh_x^6\ll1$ is deep in the low density regime but, according to condition (\ref{hopir}), the hopping remains irrelevant at most up to $\tau_Qh_x^6\approx h_x^2\approx 0.03$. To the left of this point the plots collapse to a single curve that tends to a line with the predicted slope $\frac13(81/64)^2\pi=1.67$ (see the inset). To the right the plots begin to diverge demonstrating the breakdown of the simple LZ theory.
}
\label{fig:hopping} 
\end{figure}

In order to demonstrate the interplay between the nucleation of $3$-bubbles and their hopping we simulate a ramp from $h_z^{\rm in}=-6$ to $h_z^{\rm fin}=0.8$. The density of $3$-bubbles is shown in Fig.~\ref{fig:hopping} as a function of the scaling variable deep in the low density regime, where $\tau_Q h_x^6\ll1$. With increasing $\tau_Q h_x^6$ there is a crossover from the pure LZ nucleation to the regime where the hopping becomes relevant. In the former we can see good agreement with the LZ theory, demonstrated by the collapse, while in the latter the curves begin to diverge slowly.

\textbf{Beyond low density.--} 
Upto now, we have seen that bubble nucleations at low densities are accurately described by two-level LZ problems. Moreover, for $n=1,2,3$, $\tau_Qh_x^{2n}$ is the scaling variable when the hopping is not relevant. The next natural questions are $(1)$ if it remains such beyond the low density regime, and $(2)$ whether we can also treat bubble nucleations at high densities as LZ transitions. To answer these questions, we consider again the nucleation of $1$-bubbles near $h_z=2$ but this time in full range of $\tau_Q h_x^2$. In order to isolate the $1$-bubble nucleation in full TDVP simulations, we have to keep irrelevant not only the hopping (\ref{hopir}) but also the $3$- and $2$-bubble nucleation at $h_z=2/3,1$, respectively. This requires very small $h_x^2$ that makes $\tau_Q$ rather long, making TDVP intractable for $\tau_Q h_x^2 \gg 1$.

In order to get perfect isolation and additionally get some analytical insights, first we consider an effective Hamiltonian by projecting the original Hamiltonian \eqref{hamiltonian} into the $1$-bubble subspace. On a periodic chain of $N$ sites the subspace is spanned by the initial metastable state $\ket{0} = \ket{\psi_{\text{in}}}$ in \eqref{psiin}, the translationally invariant (TI) one $1$-bubble state, $\ket{1} = \frac{1}{\sqrt{N}}\sum_j \ket{\downarrow\dots\downarrow\uparrow_j\downarrow\dots\downarrow}$, TI two $1$-bubble state $\ket{2}$, upto the TI state with $N/2$ $1$-bubbles $\ket{N/2} = \frac{1}{\sqrt{2}}(\ket{\downarrow \uparrow \downarrow \uparrow\dots}+\ket{\uparrow \downarrow \uparrow \downarrow \dots})$. It turns out that the resulting $(N/2+1)$-dimensional effective Hamiltonian can be constructed iteratively, see \cite{supple}, and for our purpose we can consider up to $N=44$ using a standard 64-bit machine.  

Similarly as (\ref{heff_1bub}), the resulting Hamiltonian is a linear combination of two terms \cite{supple}:
\be 
H_{\rm eff} = \tilde{E}_0 +\frac{t}{\tau_Q} H_z + h_x H_x,
\label{Heff}
\ee
with $\tilde{E}_0 = \braket{0|H(h_z=2, h_x=0)|0}$. This structure allows us to rewrite the Schr\"odinger equation, $i\frac{d\ket{\psi}}{dt}=H_{\rm eff}\ket{\psi}$, as $ i\frac{d}{dt'}\ket{\psi'}= \left(\frac{t'}{\tau_Q h_x^2} H_z+H_x\right) \ket{\psi'}$.
Here $t'=h_x t$ and $\tilde E_0$ was absorbed in the phase of $\ket{\psi'}$. This demonstrates that the final density of $1$-bubbles must depend on $\tau_Qh_x^2$ as a single scaling variable. 

\begin{figure}
\includegraphics[width=0.999\linewidth]{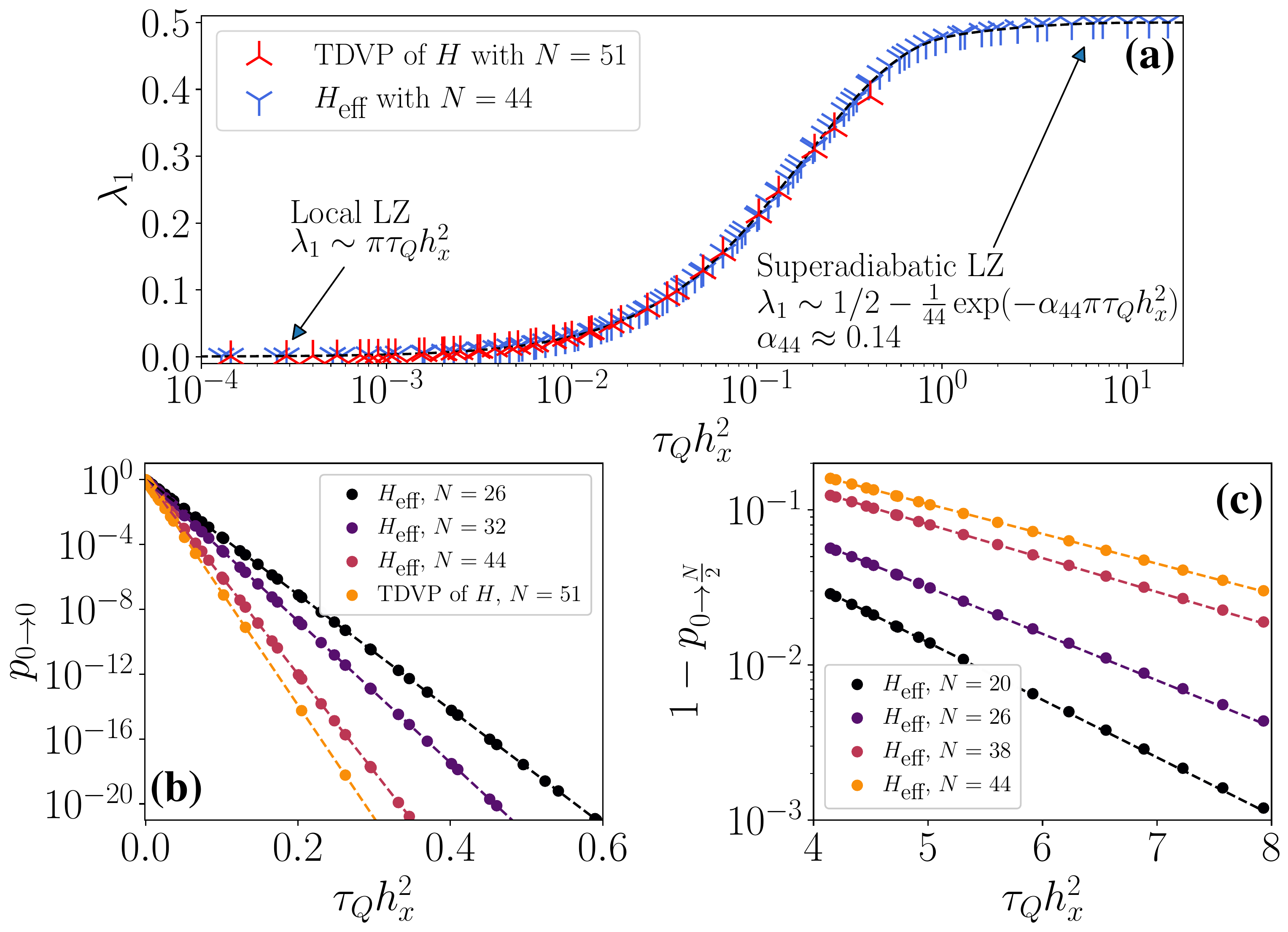}
\caption{
(a) Density of $1$-bubbles $\lambda_1$ as a function of $\tau_Q h_x^2$ after a ramp across $h_z=2$. Here the effective Hamiltonian (\ref{Heff}) is benchmarked against direct TDVP simulation. The curve crosses over from the low density linearized LZ formula (\ref{linLZ}) with $n=1$ to the high density LZ profile (\ref{highLZ}). 
(b) The multi-level LZ transition probability $p_{0 \rightarrow 0}$ as as function of $\tau_Q h_x^2$ for different system sizes $N$. The dashed line is the analytical formula: $p_{0 \rightarrow 0} = \exp\left[- N \pi \tau_Q h_x^2\right]$.
(c) The superadiabatic transition probability $p_{0 \rightarrow \frac{N}{2}}$ as a function of $\tau_Q h_x^2$ for different $N$ fitted with the exponential function $p_{0 \rightarrow \frac{N}{2}} = 1 - \exp\left[-\alpha_N \pi \tau_Q h_x^2 \right]$.
}
\label{fig:full_profile1} 
\end{figure}

The dependence, obtained by simulation with the effective Hamiltonian, is plotted in Fig. \ref{fig:full_profile1}(a). The same figure compares results from full TDVP simulations. Unlike the low density regime, the generic problem now is that of a $(N/2+1)$-level LZ transition, which again in the low density regime reduces to the local two level LZ scenario. To describe such a multi-level LZ problem, we consider two transition probabilities:
\bea
p_{0 \rightarrow 0} &=& \lim_{t \rightarrow \infty} |\braket{0 | \psi(t)}|^2,  \nonumber \\
p_{0 \rightarrow \frac{N}{2}} &=& \lim_{t \rightarrow \infty} |\braket{N/2 | \psi(t)}|^2.
\eea
Following Ref. \cite{Shytov2004}, the former one has the exact form: 
\be
p_{0 \rightarrow 0} = \exp\left[-2 \pi \tau_Q h_x^2 \sum_{n=1}^{N/2} \frac{|\braket{n|H_x|0}|^2}{|\braket{n|H_z|n} - \braket{0|H_z|0}|}\right],
\ee
which translates into $p_{0 \rightarrow 0} = \exp\left[- N \pi \tau_Q h_x^2\right]$~\cite{supple}. Fig. \ref{fig:full_profile1}(b) shows the profile of $p_{0 \rightarrow 0}$ for different values of $N$ that perfectly matches the analytical prediction. Moreover, for $\tau_Q h_x^2 \ll 1$ only transitions between $\ket{0}$ and states $\ket{n}$ with low density, $n\ll N$, become relevant. The total probability of these transitions is $1 - p_{0 \rightarrow 0}$. Therefore, in this regime the density of 1-bubbles $\lambda_1 = \frac{1}{N}(1 - p_{0 \rightarrow 0}) \approx \pi \tau_Q h_x^2$ confirming the earlier analysis again.

On the other hand, when the curve in Fig. \ref{fig:full_profile1}(a) reaches the superadiabatic regime, $\tau_Q h_x^2 \gg 1$, there is only one relevant LZ anticrossing. The initial metastable state $\ket{0}$ crosses over to the final state $\ket{N/2}$ with probability $p_{0 \rightarrow \frac{N}{2}}$ and $(1 - p_{0 \rightarrow \frac{N}{2}})$ becomes a small excitation probability to the state $\ket{N/2-1}$. An analytical derivation of $p_{0 \rightarrow \frac{N}{2}}$ from the multi-level LZ problem is beyond the scope this work. However, we find the following form
\be
p_{0 \rightarrow \frac{N}{2}} = 1 - \exp\left[-\alpha_N \pi \tau_Q h_x^2 \right],
\ee
where the coefficient $\alpha_N$ decreases with $N$~\cite{supple}, see Fig. \ref{fig:full_profile1}(c). 
Therefore, the $1$-bubble density in this regime is
\bea 
\lambda_1 &=&
\frac{1}{N}
\left[
\frac{N}{2} p_{0 \rightarrow \frac{N}{2}} +
\left(\frac{N}{2}-1\right) (1-p_{0 \rightarrow \frac{N}{2}})
\right] \nonumber \\
&=& \frac12-\frac{1}{N} (1 - p_{0 \rightarrow \frac{N}{2}}),
\label{highLZ}
\eea 
which is in good agreement with Fig. \ref{fig:full_profile1}(a).

\textbf{Conclusion and outlook.--} 
We have shown that the metastability pertained to FOQPT in the quantum Ising model under transverse and longitudinal fields is lost in successive stages in quenches across the FOQPT point, that occurs due to \textit{quantized} the nucleation of bubbles. Specifically, we have identified special resonant regions in the longitudinal field ($h_z = 2/n$), where the metastable state can easily tunnel to nucleate bubbles of specific size $n$, which are $n$-th order perturbative processes in the transverse field $h_x$.
Moreover, we have unified this entire non-adiabatic process under the umbrella of Landau-Zener theories -- 
the low density nucleations can be understood through two-level Landau-Zener transitions, while at higher densities the situations translate to the multi-level Landau-Zener problems.

Furthermore, our work can be easily generalized to higher dimensions, where the special resonant points become
$
h_z \propto S/V.
$
Here $S$ is the surface area and $V$ the volume of a bubble, each of them taking discrete values. The physical implementation of the transverse Ising model with a chain of Rydberg atoms provided spectacular demonstration~\cite{Lukin18} of the quantum Kibble-Zurek mechanism. Within two years following this breakthrough, the number of Rydberg atoms increased from $50$ in 1D \cite{Lukin18} to a few hundred in 2D/3D structures \cite{rydberg2d1,rydberg2d2}. Such marvelous achievements on the experimental front make possible to explore regimes where the nucleation of bubbles manifests a quantized nature, not only in 1D but also in higher dimensions.

\acknowledgements
We are grateful to Marek M. Rams, Debasis Sadhukhan, Jakub Zakrzewski,  Maciej Lewenstein, and Luca Tagliacozzo for useful discussions and valuable comments.
We acknowledge funding by the National Science Centre (NCN), Poland together with European Union through 
QuantERA ERA NET programs: NAQUAS 2017/25/Z/ST2/03028 (AS, JD) and QTFLAG 2017/25/Z/ST2/03029 (TC).

\bibliography{first_order.bbl}

\end{document}


\title{Supplementary material to ``Non-adiabatic dynamics across a first order quantum phase transition:
Quantized bubble nucleation''}

\author{Aritra Sinha}
\email{aritrasinha98@gmail.com}
\affiliation{Institute of Theoretical Physics, Jagiellonian University, \L{}ojasiewicza 11, 30-348 Krak\'ow, Poland }

\author{Titas Chanda}
\email{titas.chanda@uj.edu.pl}
\affiliation{Institute of Theoretical Physics, Jagiellonian University, \L{}ojasiewicza 11, 30-348 Krak\'ow, Poland }

\author{Jacek Dziarmaga}
\email{dziarmaga@th.if.uj.edu.pl}
\affiliation{Institute of Theoretical Physics, Jagiellonian University, \L{}ojasiewicza 11, 30-348 Krak\'ow, Poland }

\date{\today}

\begin{abstract}
Here we present the derivation of effective two-level Hamiltonians near the resonant points using Schrieffer-Wolff transformation and the general Landau-Zener probability for an $n$-bubble nucleation process. 
Furthermore, we present
some additional details to complement the main text.
Specifically, we discuss details about the multi-level Landau-Zener problem beyond the low density regime.
\end{abstract}

\maketitle

 
\subsection{Effective Hamiltonian near second order resonant point} 

To understand the physics near the $2^{\text{nd}}$ order resonant point (RP${}_2$) at $h_{z}=1$, we need to construct a simple effective Hamiltonian using the Schrieffer-Wolff transformation~\cite{Wolff1966}. 
Following the idea of low density approximations discussed in the main text
only the following \textit{local} process is allowed near RP${}_2$:
\be 
\underset{\ket{M}}{\ket{\downarrow\dots\downarrow \downarrow_j\downarrow_{j+1} \downarrow\dots\downarrow }}
\stackrel{h_x^2}{\longleftrightarrow} 
\underset{\ket{T_{2}}}{\ket{ \downarrow\dots\downarrow \uparrow_j\uparrow_{j+1} \downarrow\dots\downarrow }}.
\ee 
The calculation done below closely follows similar calculations done in references ~\cite{Olexei2017,Heyl2020}.  We take $H = H^{0}+V$,
with
\bea 
H^{0} = -\sum_{n=1}^N \left[\sigma^z_n\sigma^z_{n+1}  + h_{z}\sigma^z_n\right]
\label{solvable}
\eea 
being the unperturbed Hamiltonian and 
\bea 
V = -h_{x}\sigma^x_n
\label{pert}
\eea
as the perturbation. 

We carry out a unitary transformation with generator $S$ such that the rotated Hamiltonian has no first order order term:
\bea
\Hat{\mathcal{H}} &=& e^{S}He^{-S} \nonumber \\
          &=& H^{0}+V+[S,H^{0}]+[S,V]+\frac{1}{2}[S,[S,H^{0}]] + \ldots . \nonumber \\
\label{sw_trans}
\eea
Therefore, we need to set $[S,H^{0}] = -V$
to get rid of first order perturbation, and in  the basis of the unperturbed Hamiltonian $H^{0}$, we get 
\bea 
S_{ij} = V_{ij}/(\epsilon_{i}-\epsilon_{j})
\label{S_mat}
\eea
where $V_{ij} = \braket{i|V|j}$, providing us our rotated Hamiltonian 
\bea 
\Hat{\mathcal{H}} = H^{0}+\frac{1}{2}[S,V]
\label{comp_ham}
\eea
up to second order in perturbation.

\begin{figure}[t]
    \centering
    \includegraphics[width=\linewidth]{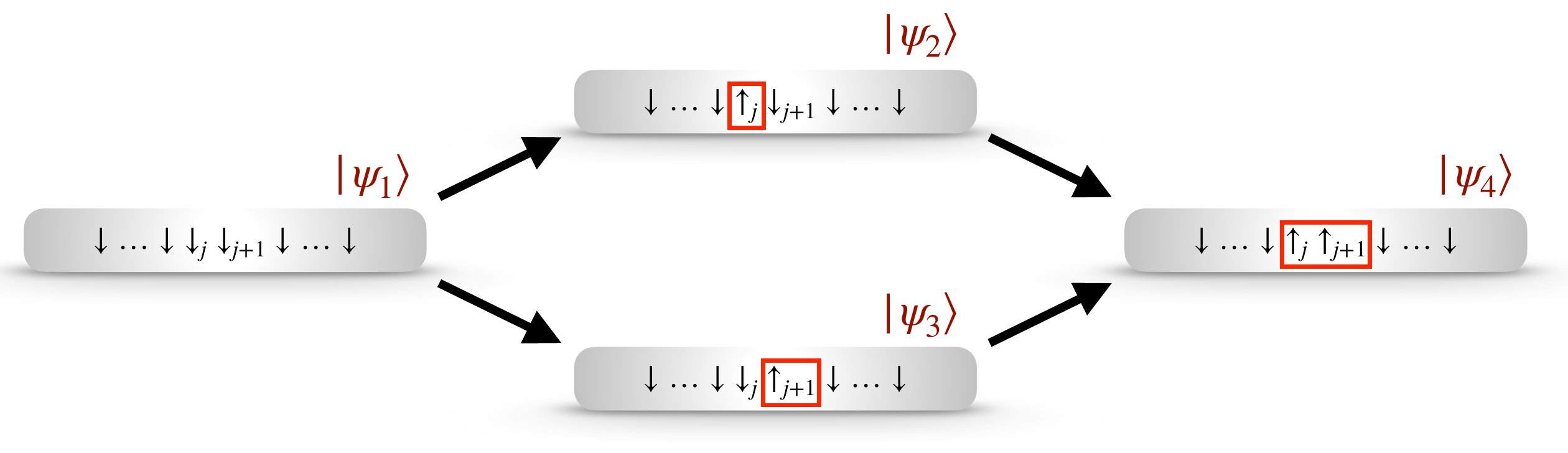}
    \caption{The sketch illustrates the the Ising states relevant in the process connecting the meta-stable state $\ket{\psi_{1}}$ to the target $2-$bubble nucleated state $\ket{\psi_{4}}$ under the effective second order (in perturbation) Hamiltonian $H_{\text{eff}}^{2}$. Each dark arrow connects two Ising states by an order of perturbation $h_{x}$.}
    \label{schematic_2}
\end{figure}

Now, we need to identify the shortest paths which connect the states $\ket{M}$ and $\ket{T_{2}}$. This takes place through the states $\ket{C_{1}} = \ket{\downarrow ... \downarrow \uparrow_{j} \downarrow_{j+1} \downarrow  ... \downarrow}$ and $\ket{C_{2}} = \ket{\downarrow ...  \downarrow \downarrow_{j} \uparrow_{j+1}\downarrow ... \downarrow}$.
Thereafter we construct the relevant subspace using the four states (see Fig.~\ref{schematic_2} for the processes involving the states in the subspace): 
$$
\ket{\psi_{1}} = \ket{M}; \ket{\psi_{2}} = \ket{C_{1}}; \ket{\psi_{3}} = \ket{C_{2}}; \ket{\psi_{4}}= \ket{T}.
$$
In this subspace, we have the unperturbed Hamiltonian $H^{0}$ as a $4 \times 4 $ diagonal matrix containing the non-zero elements: $H^{0}_{ii}= \bra{\psi_{i}}H^{0}\ket{\psi_{i}}$. Similarly from~\eqref{pert} and ~\eqref{S_mat}, we get 
\begin{equation}
V = -h_{x}  \begin{bmatrix}
0 & 1 & 1 & 0\\
1 & 0 & 0 & 1 \\
1 & 0 & 0 & 1\\
0 & 1 & 1 & 0
\end{bmatrix} \hspace*{0.2cm}\text{and} \hspace*{0.2cm} S = \begin{bmatrix}
0 & a & a & 0\\
-a & 0 & 0 & b \\
-a & 0 & 0 & b\\
0 & -b & -b & 0
\end{bmatrix}
\end{equation}
where $a = \frac{h_{x}}{2h_{z}-4}$ and $b = -\frac{h_{x}}{2h_{z}}$.

Now if we take into account only the basis spanned by the initial meta-stable state($\ket{\psi_{1}}$) and the target resonant state($\ket{\psi_{4}}$), we get an effective two-level system as desired for Landau Zener probability calculations:
\begin{align*}
H^{(2)}_{\text{eff}} =  \begin{bmatrix}
H^{0}_{11}-\frac{h_{x}^2}{2-h_{z}} & & -\frac{h_{x}^2}{2h_{z}-h_{z}^2}\\
-\frac{h_{x}^2}{2h_{z}-h_{z}^2} & & H^{0}_{44}-\frac{h_{x}^2}{h_{z}}
\label{eff_hamm}
\end{bmatrix}
\end{align*}
where $H^{0}_{11} = -(N-1)+N h_{z}$ and $H^{0}_{44} = -(N-5)+(N-4)h_{z}$. Near RP${}_{2}$  at $h_{z} = 1$, the off-diagonal terms reduce to just $-h_{x}^2$; while in the limit $h_{x}\ll 1$, the $h_{x}^2$ contribution in the diagonal terms can be neglected. The final form is 
\begin{equation}
H_{\text{eff}}^{(2)} \approx E_0(h_z)+\begin{bmatrix}
0 & -h_{x}^{2}\\
-h_{x}^{2} & 4-4h_{z} 
\end{bmatrix}.    
\label{SW2_eff}
\end{equation}
Here $E_0(h_z)=H^{0}_{11}=-(N-1)+Nh_z$ is energy of the meta-stable state.
 

\subsection{Effective Hamiltonian near third order resonant point} 

\begin{figure*}
    \includegraphics[width=0.85\linewidth]{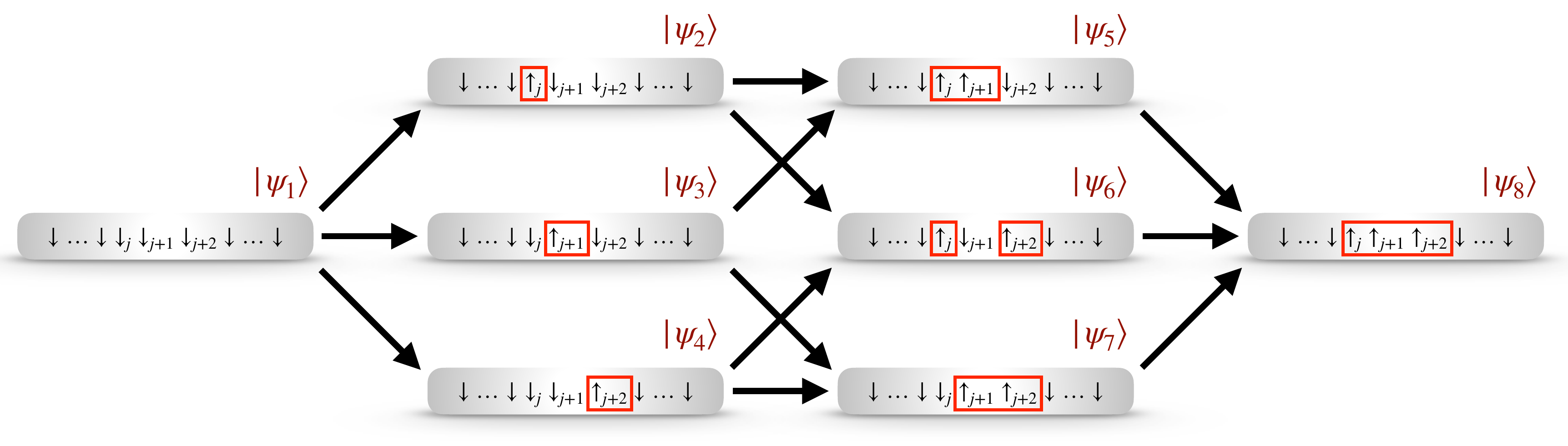}
    \caption{The sketch illustrates the the Ising states relevant in the process connecting the meta-stable state $\ket{\psi_{1}}$ to the target $3-$bubble nucleated state $\ket{\psi_{8}}$ under the effective $3^{\text{rd}}$ order (in perturbation) Hamiltonian $H_{\text{eff}}^{3}$. Each dark arrow connects two Ising states by an order of perturbation $h_{x}$.}
    \label{schematic_3}
\end{figure*}

The processes allowed near the $3^{\text{rd}}$ resonant point (RP${}_{3}$) at $h_{z} = 2/3$ involve a $3$-bubble nucleation as the following:
\be 
\underset{\ket{M}}{\ket{\downarrow\dots\downarrow \downarrow_j\downarrow_{j+1}\downarrow_{j+2} \downarrow\dots\downarrow }}
\stackrel{h_x^3}{\longleftrightarrow} 
\underset{\ket{T_{3}}}{\ket{ \downarrow\dots\downarrow \uparrow_j\uparrow_{j+1}\uparrow_{j+2} \downarrow\dots\downarrow }},
\ee  
Here too we transform the Hamiltonian through a generator $S$, to get rid of the first order and second order terms and put into prominence the third order perturbation which governs the low energy physics near RP${}_{3}$. To do that we expand $S = S_{1}+S_{2}$ as a sum of two terms. Here $S_{1} \sim \mathcal{O}(h_{x})$ kills the first order terms of the transformed Hamiltonian and $S_{2} \sim \mathcal{O}(h_{x}^{2})$ gets rid of the second order terms. The transformation applied on the Hamiltonian goes like $\Hat{\mathcal{H}} = e^{S_{2}}e^{S_{1}}He^{-S_{1}}e^{-S_{2}}$. As before we collect the first order and second order terms and set them to $0$ to obtain two conditions in order to determine the explicit forms of $S_{1}$ and $S_{2}$:
\begin{align}
  [S_{1},H_{0}]&=-V,\nonumber\\
  [S_{2},H_{0}]&=-\frac{1}{2}[S_{1},V].
  \label{cond_S2}
\end{align}

Using~\eqref{cond_S2}, the transformed Hamiltonian upto $3^{\text{rd}}$ order in perturbation looks like :
\be
\Hat{\mathcal{H}} = H_{0}+[S_{2},V] + \frac{1}{2}[S_{1}H_{0}S_{1},S_{1}]+\frac{1}{6}[S_{1}^{3},H_{0}].
\label{SW3rd_eff_full}
\ee

Having done the mathematical analysis, we need to construct the basis which connects $\ket{M}$ to $\ket{T_{3}}$. See the sketch in Fig.~\ref{schematic_3} as an aid to visualize the perturbative processes involved among the states in the subspace, and also for the labels of states considered in the following discussion. There are a minimum $8$ unique basis states in such a process: $\{\ket{\psi_{i}}$; $i = 1, \ldots,8\}$ with $\ket{\psi_{1}} = \ket{M}$ and $\ket{\psi_{8}} = \ket{T_{3}}$. After performing calculations with the basis states, we arrive at an effective two-level form of the rotated Hamiltonian in~\eqref{SW3rd_eff_full} connecting  $\ket{\psi_{1}}$ and  $\ket{\psi_{8}}$ through a third order process at $h_{z}=2/3$:

\begin{equation}
\label{SW3_eff}
H_{\text{eff}}^{(3)} \approx E_0(h_z)+\begin{bmatrix}
0 & -\frac{81}{64} h_{x}^{3}\\
-\frac{81}{64}h_{x}^{3} & 4-6h_{z} 
\end{bmatrix}.    
\end{equation}


\subsection{Landau Zener Calculations} 

In this section of the supplementary text, we calculate the Landau Zener probability to nucleate a bubble of size $n$ near $h_z=2/n$, from an initial metastable state. To do this we apply the Landau-Zener formula~\cite{Landau65, Zener32, Stueckelberg32, Majorana32} after transforming the effective two-level problem to an appropriate framework. 

Consider flipping $n$ consecutive spins $j,\dots,j+n-1$ by a $n$-th order process:

\be 
\underset{\ket{A}}{\ket{ \downarrow\dots\downarrow \downarrow_j\dots\downarrow_{j+n-1} \downarrow\dots\downarrow }} 
\stackrel{h_x^n}{\longleftrightarrow} 
\underset{\ket{B}}{\ket{ \downarrow\dots\downarrow \uparrow_j  \dots  \uparrow_{j+n-1} \downarrow\dots\downarrow }}.
\ee 
For such a process the general effective Hamiltonian reads
\begin{equation}
H_{\text{eff}}^{(n)} \approx E_0(h_z)+
\begin{bmatrix}
0          & -c_n h_{x}^{n}\\
-c_n h_{x}^{n} & 4-2n h_{z} 
\label{heff_nbub_supp1}
\end{bmatrix}.    
\end{equation} 
where $c_{n}$ is a constant for a specific perturbative process. Comparing with effective Hamiltonians~\eqref{SW2_eff} and ~\eqref{SW3_eff} in the previous sub-sections, it is $c_{2} = 1$ for a second order and $c_{3} = 81/64$ for third-order process.

Rescaling the energy of effective Hamiltonian for such processes (see~\eqref{heff_nbub_supp1}), we arrive at 
\begin{align*}
H_{\text{eff}}^{(n)} \approx 
\begin{bmatrix}
-2+ n h_{z}         & -c_n h_{x}^{n}\\
-c_n h_{x}^{n} & 2-n h_{z} 
\label{heff_nbub_supp2}
\end{bmatrix}.    
\end{align*} 
Let us linearize the longitudinal field as 
\bea 
h_{z} = t/\tau_{Q} +2/n
\label{quench_profile_hz_2}
\eea 
where $t$ goes from $-\infty$ to $\infty$ such that at $t = 0$ the ramp is centered on the resonant point at $h_{z} = 2/n$. If 
\be
\ket{\psi(t)} = \alpha(t) \ket{A} + \beta(t) \ket{B}
\ee
 then the Schrodinger equation $i\frac{d}{dt}\ket{\psi} = H_{\text{eff}}^{(n)
 }\ket{\psi}$ for $\alpha(t)$ and $\beta(t)$ is given by a set of coupled differential equations:
\be
i\frac{d}{dt}\left(\begin{array}{c}  \alpha\\
\beta
\end{array}
\right)=  \left[
\frac{nt}{\tau_Q}\sigma^z-c_{n}h_{x}^{n}\sigma^x
\right]
\left(
\begin{array}{c}
\alpha \\
\beta
\end{array}
\right)
\label{tLZ_1}
\ee
Defining $t' = -nth_{x}$ and $\Delta = {(2\frac{c_{n}^{2}}{n}\tau_{Q}h_{x}^{2n})}^{-1}$ we get
\be
i\frac{d}{dt'}\left(\begin{array}{c}  \alpha\\
\beta
\end{array}
\right)=  \frac12 \left[
\Delta t'\sigma^z+\sigma^x
\right]
\left(
\begin{array}{c}
\alpha \\
\beta
\end{array}
\right)
\label{tLZ_2}
\ee
Say we initialize the system in state $\ket{A}$. The initial condition is then given by $\alpha(t=-\infty) = 1, \hspace*{0.3cm} \beta(t=-\infty) = 0$. Here we are interested in the occupation probabilities of the states at $t\rightarrow +\infty$. Note that after the anti-crossing ($t>0$) the metastable state $\ket{A}$ is the excited state and if the transition in the parent model was fully adiabatic the probability to stay in $\ket{B}$ at $t\rightarrow +\infty$ would be $1$. With this consideration, the probability for the system to remain in $\ket{B}$ or in the present context, the probability that a spin flip of size $n$ occurs at the chosen site in $\ket{B}$ is given by $|\beta(t=\infty)|^{2} = 1-\exp(-\pi/2\Delta) \approx  \pi/2\Delta = \frac{c_{n}^{2}}{n}\pi\tau_{Q}h_{x}^{2n}$ assuming $\tau_{Q}h_{x}^{2n} \ll 1$ for physical reasons elaborated in the main text.


 \subsection{Calculation of the observables}
 
\begin{figure}[t]
    \includegraphics[width=\linewidth]{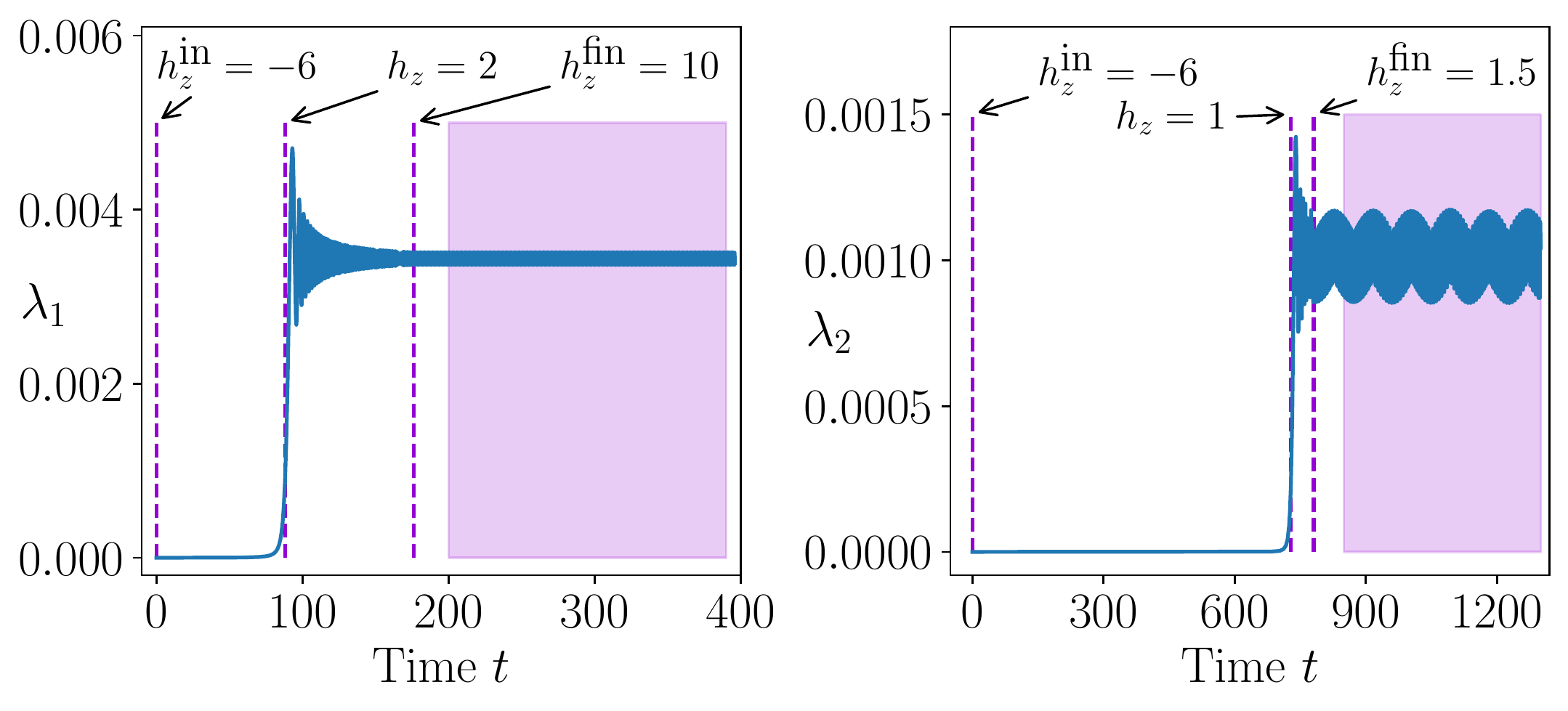}
    \caption{(Left) Density profile of $1-$bubbles $\lambda_{1}$ versus time for a fixed transverse field strength $h_x = 0.01$ for $\tau_Q = 11$. (Right) Profile of the $2-$bubble density $\lambda_{2}$ with time for transverse field strength $h_x = 0.05$ with $\tau_Q=104$. In both the cases, we stop the ramp at $h_{z}^{\rm fin}$ and then perform free evolution with fixed $h_z^{\rm fin}$. The averaging is performed over the oscillations in the region shaded with violet.}
    \label{app_obs}
\end{figure}

 To verify the predictions offered by Landau-Zener theory we calculate the $n$-bubble density (Eq.(\textcolor{blue}{10}) in the main text). In that respect, we follow the quench protocol of Eq. (\textcolor{blue}{2}) of the main text upto $h_z = h_z^{\text{fin}}$, then proceed with free evolution with the fixed longitudinal field $h_z^{\text{fin}}$ for a long time.
 In the case of $1$-bubbles, the ramp is simulated from $h_z^{\rm in}=-6$ to $h_z^{\rm fin}=10$, while for $2$-bubbles we set $h_{z}^{\rm fin}$ to be $1.5$, and for $3$-bubbles $h_{z}^{\rm fin} = 0.8$. In order to get $n$-bubble density $\lambda_n$ ($n = 1, 2, 3$), we average generously over the oscillations during the free evolution. See Fig.~\ref{app_obs} for sample time profiles for the observables $\lambda_{1}$ and $\lambda_{2}$ during the ramp. We denote the  $h_z^{\rm in}$ and $h_z^{\rm fin}$, i.e., the start and end-points of the ramp by dashed vertical lines and the violet shaded region demarcates the region of the oscillatory part we average over.

\subsection{Beyond the low density regime: details}

 To look at the $1$-bubble nucleation in full isolation, we project our original Hamiltonian (Eq. (\textcolor{blue}{$1$}) in the main text) onto a $1$-bubbles subspace. We start with the initial metastable state $\ket{0} = \ket{\psi_{\textrm{in}}}$ (Eq. (\textcolor{blue}{$3$}) in the main text), and generate states with increasing number of $1$-bubbles iteratively from that. For example, single spin flips on $\ket{0}$ will create the translationally invariant (TI) one $1$-bubble state
 \be 
\ket{1} = \frac{1}{\sqrt{N}}\sum_j \ket{\downarrow\dots\downarrow\uparrow_j\downarrow\dots\downarrow},
\ee
another set of single spin flips on $\ket{1}$ generates the TI two $1$-bubble state
\be  
\ket{2} = \frac{1}{\sqrt{N(N-1))}}\sum_{\substack{j_{1},j_{2} \\ j_{2}
\neq j_{1}\pm 1}} \ket{\downarrow\dots\downarrow\uparrow_{j_{1}}\downarrow\dots\downarrow\uparrow_{j_{2}}\downarrow\dots \downarrow},
\ee  
upto the TI $N/2$ $1$-bubble state.
\be
\ket{N/2} = \frac{1}{\sqrt{2}}(\ket{\downarrow \uparrow \downarrow \uparrow\dots}+\ket{\uparrow \downarrow \uparrow \downarrow \dots})
\ee
Moreover, in this scheme, the $n$ $1$-bubble state can be constructed iteratively from $(n-1)$ $1$-bubble state by applying $\sum_{j}\sigma^{+}_{j}$ on the state $\ket{n-1}$ with the \textit{condition that no neighboring spins can be $\uparrow$-spin}, i.e., 
 \be   
 \ket{n-1} \stackrel{\sum_{j}\sigma_{j}^{+} \text{ with the condition }}{\xrightarrow{\hspace*{3cm}}} \frac{D_{n}}{\sqrt{F_{n-1}F_{n}}}\ket{n},
 \ee 
 where $D_{n}$ and $F_{n}$s are combinatorial integers. 
 The projected Hamiltonian in the basis $\{\ket{n}, n=0,1,...,N/2 \}$ has the following tri-diagonal form 
 \bea
({H}_{\text{eff}})_{(n,n)} &=& 2n(2-h_{z})+ E_0(h_z), \nonumber \\
(H_{\text{eff}})_{(n,n-1)} &=& (H_{\text{eff}})_{(n-1,n)} = -h_{x}\frac{D_{n}}{\sqrt{F_{n-1}F_{n}}}.
\eea 
 Therefore, following the convention of Eq. (\textcolor{blue}{$12$}) in the main text, we get
 \bea
(H_z)_{(n, n)} &=& -2 n, \nonumber \\
(H_x)_{(n, n-1)} &=& (H_x)_{(n-1, n)} = -\frac{D_{n}}{\sqrt{F_{n-1}F_{n}}}.
\eea

To get the transition probability $p_{0 \rightarrow 0}$ (see Eq. (\textcolor{blue}{$14$}) of the main text), only relevant terms are $(H_x)_{(1, 0)}$ and $(H_z)_{(0, 0)}$. Moreover, it is easy to check that  $(H_x)_{(1, 0)} = (H_x)_{(0, 1)} = \sqrt{N}$, giving us the closed analytical form $p_{0 \rightarrow 0} = \exp\left[- N \pi \tau_Q h_x^2\right]$.

Finding the closed analytical form of the transition probability $p_{0 \rightarrow \frac{N}{2}}$ is beyond the scope of our work, and we resort to numerical results. As already mentioned in the main text, we find that
\be
p_{0 \rightarrow \frac{N}{2}} = 1 - \exp\left[-\alpha_N \pi \tau_Q h_x^2 \right],
\ee
with the coefficient $\alpha_N$ being dependant on the system size $N$. Below, we tabulate the numerical values of $\alpha_N$ found for different system-sizes:
\begin{center}
\begin{tabular}{|c|c|c|c|c|c|c|}
\hline
System size $N$ & \ 14 \ & \ 20 \ & \ 26 \ & \ 32 \ & \ 38 \ & \ 44 \ \\
\hline
Coefficient $\alpha_N$ & 0.36 & 0.27 & 0.22 & 0.19 & 0.16 & 0.14 \\
\hline
\end{tabular}
\end{center}
Here the errors in the fitting are below $10^{-4}$.

\bibliography{supple_first_order.bbl}
